\documentclass[12pt]{revtex4-1}
\usepackage{graphicx}

\pdfoutput=1

\begin{document} 

% title and authors

\title{Comment on ``Observing the Average Trajectories of Single Photons in a Two-Slit Interferometer''} 

\author{Timothy M. Coffey}
\author{Robert E. Wyatt}
\affiliation{Department of Chemistry and Biochemistry and Institute for Theoretical Chemistry\\
1 University Station A5300, University of Texas at Austin, Austin, TX 78712, USA}
\email[Correspondence: ]{tcoffey@physics.utexas.edu}
\date{\today}

\begin{abstract}
Kocsis {\it et al.} (Science, Reports 3 June 2011, p. 1170) state that the experimentally deduced average photon trajectories are identical to the particle trajectories of Bohm's quantum mechanics. No supporting evidence, however, was provided. The photon trajectories presented in their report do not converge to high probability regions, a familiar and necessary behavior of Bohm trajectories. We reanalyze their data and calculations, conclude that the average photon trajectories do indeed agree with Bohm, and discuss possible interpretations of this result. 
\end{abstract}
\maketitle

A recent report, Kocsis {\it et al.} \cite{Kocsis2011}, provided an innovative application of weak measurement theory \cite{Aharonov1988}. 
From weak measurements of momentum (encoded by photon polarization) and strong measurements of position, they were able to deduce the average photon trajectory in a two-slit apparatus. Techniques, such as these, provide deeper probes into the quantum world \cite{Lundeen2011}. The authors claim that the average photon trajectories are identical to Bohm's quantum particle trajectories \cite{Holland1995,Wyatt2005}, although no evidence was offered. A well known behavior of Bohm trajectories is that they congregate in regions of high probability. Figures 3 and 4 in their report show a surprising low number of photon trajectories in regions of high probability, therefore, we contacted the authors and they graciously provided us with their data and Matlab codes. Upon examining their codes we noticed several minor errors, and that only about half of the collected data was shown in their report. We independently produced corrected program codes and recomputed the photon trajectories. Unfortunately, the photon trajectories still did not congregate in high probability regions. Next, we altered the corrected calculation by using standard density estimation techniques, and found that the resulting photon trajectories, like the Bohm trajectories, did congregate in the high probability regions. Finally, we discuss an interpretation of the average photon trajectories.

The computer codes supplied to us by Kocsis {\it et al.} contained several minor errors (see details in appendix). We corrected these errors, and performed major parts of the calculation in a separate program. This revised calculation resulted in average photon trajectories (red dots), see Fig.~1(a), that still did not congregate in high probability regions, with the largest discrepancy in the central bright band on the right hand side of the figure. In contrast, the Bohm trajectories (blue lines) are dense in accordance with the intensity bands. The inset (b) shows the range of data presented by Kocsis {\it et al.}, and many photon trajectories moving in the opposite direction of their Bohm counterparts. In Fig.~1(c) the probability density for the final $z$ step is shown, and the final position of each trajectory is superposed. Again, the lack of photon trajectories in the high probability regions is evident. The photon trajectories, however, do seem to coincide with the Bohm trajectories. We calculated Pearson's sample correlation coefficient for each photon and Bohm trajectory pair, and the mean over the ensemble was $r_{\rm avg} = 0.53$.

\begin{figure}
\centerline{\includegraphics{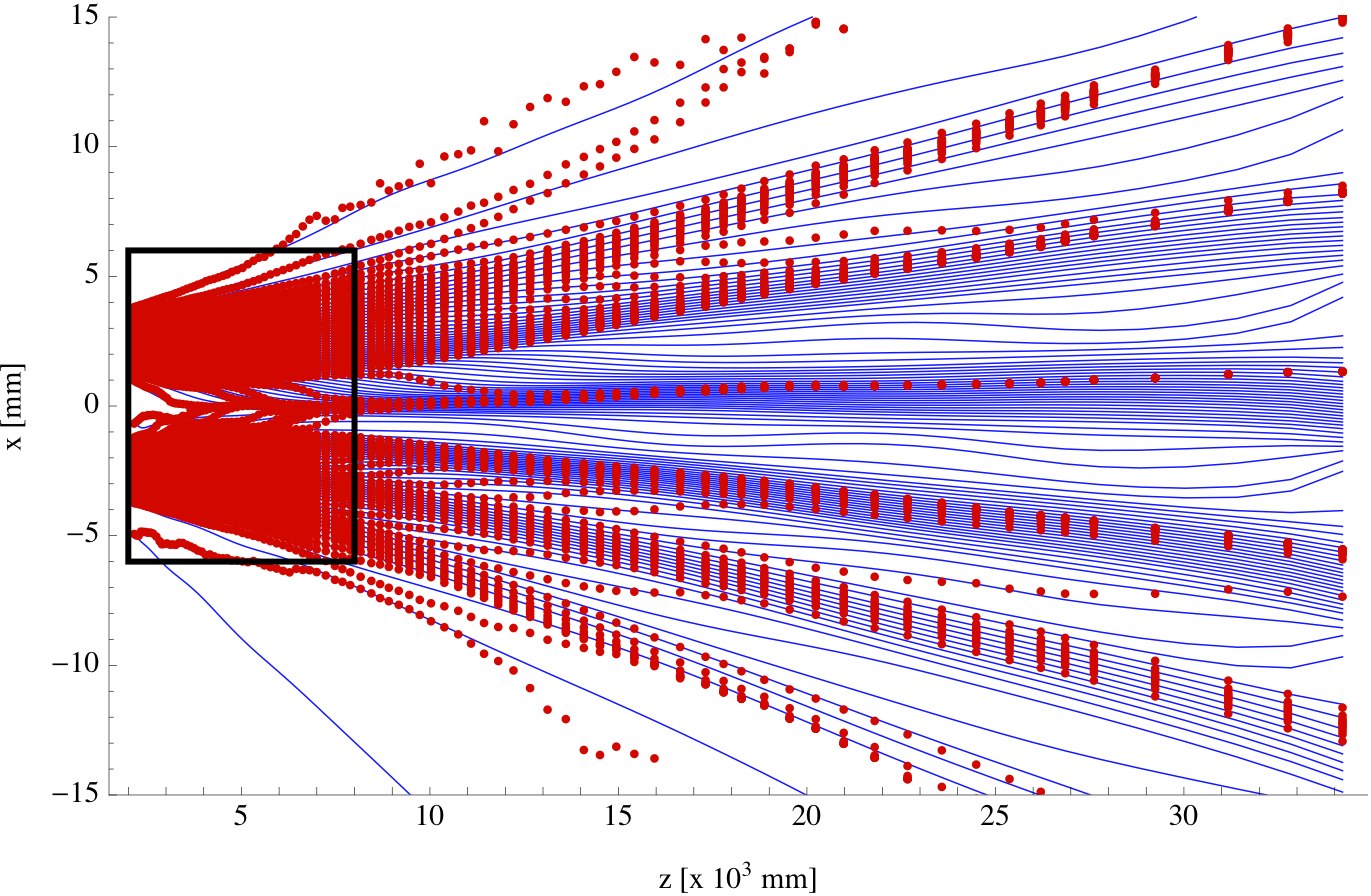}}
\centerline{(a)}
\centerline{\includegraphics{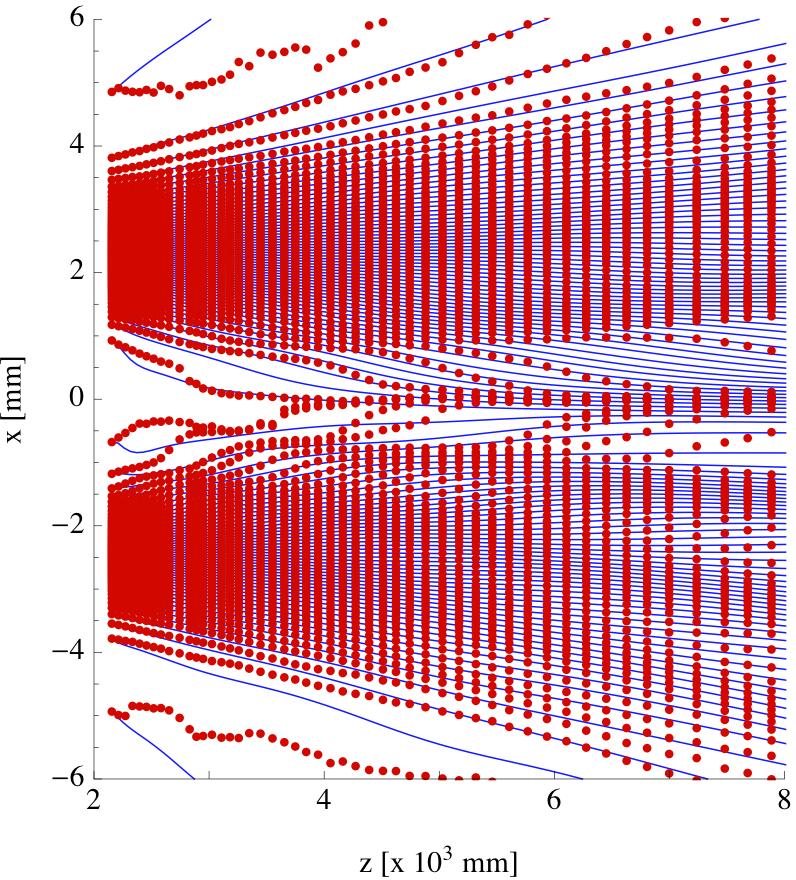}
\hspace{0.3in}\includegraphics{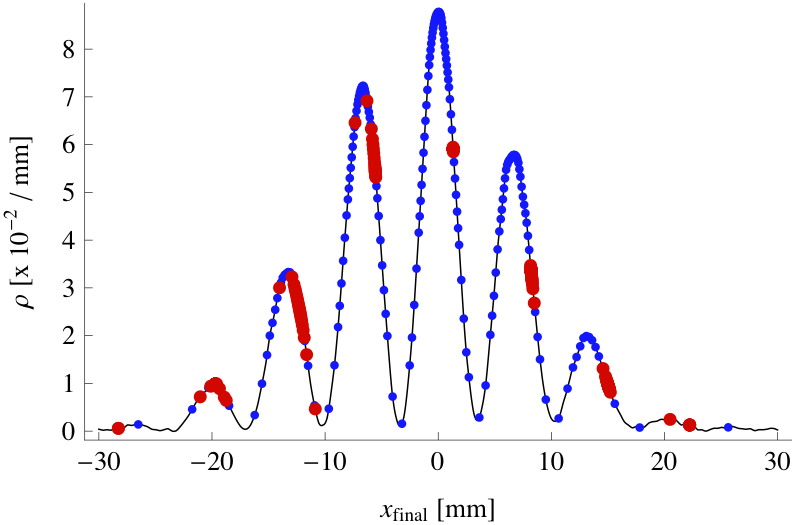}}
\centerline{(b)\hspace{3in}(c)}
\caption{Comparison of the average photon trajectories (red dots) and the Bohm trajectories (blue lines) using our corrected computer codes. The photon trajectories do not bunch correctly in the high probability regions, especially in the central bright spot. The inset (b) highlights many photon and Bohm trajectories moving oppositely. Part (c) shows the final trajectory positions as compared to the probability density. There are too few photon trajectories near the peaks of the density.}
\end{figure}

The Kocsis {\it et al.} experiment inferred the transverse or $x$ momentum of the photons by,
\begin{equation}
\frac{k_x}{|\mathbf{k}|} = 
\frac{1}{\zeta}\left[ \sin^{-1} \left(
\frac{I_{\rm R} - I_{\rm L}}{I_{\rm R} + I_{\rm L}}
\right)   \right]
\end{equation}
where the coupling coefficient $\zeta$ was found to be close to $373.5$, and $I_{\rm R}$ and $I_{\rm L}$ are the two measured intensities, corresponding to right and left circular polarizations, respectively. The intensities were measured at each $z$ step on a CCD with a pixel size of $26~\mu\rm m$. The background noise was subtracted from each pixel count. During their calculation (and our corrected one above) each intensity was approximated by cubic splines using the counts at each pixel. This resulted in jagged intensities. To improve the data fitting, we instead estimated each density using a Gaussian kernel estimator,
\begin{equation}
I_{\rm est} (x) = 
\frac{1}{n h} \sum_i^n 
K \left( \frac{x - x_i}{h} \right),
\end{equation}
where $n$ is the number of data points, $K(u) = \rm exp(-u^2 / 2 ) / \sqrt{2 \pi}$, and the bandwidth $h$ was set to the standard Silverman \cite{Silverman1986} value of $1.06 \hat\sigma n^{-1/5}$, with $\hat\sigma$ being the sample standard deviation. In Fig.~2 are shown the average photon trajectories (again in red dots) compared to the Bohm particle trajectories (blue lines). The inset shows that the average photon trajectories are now following the Bohm trajectories. The photon trajectories also now congregate better in the high probability regions. By using density estimation the mean correlation coefficient improved by 17\% to $r_{\rm avg}=0.62$.

\begin{figure}
\centerline{\includegraphics{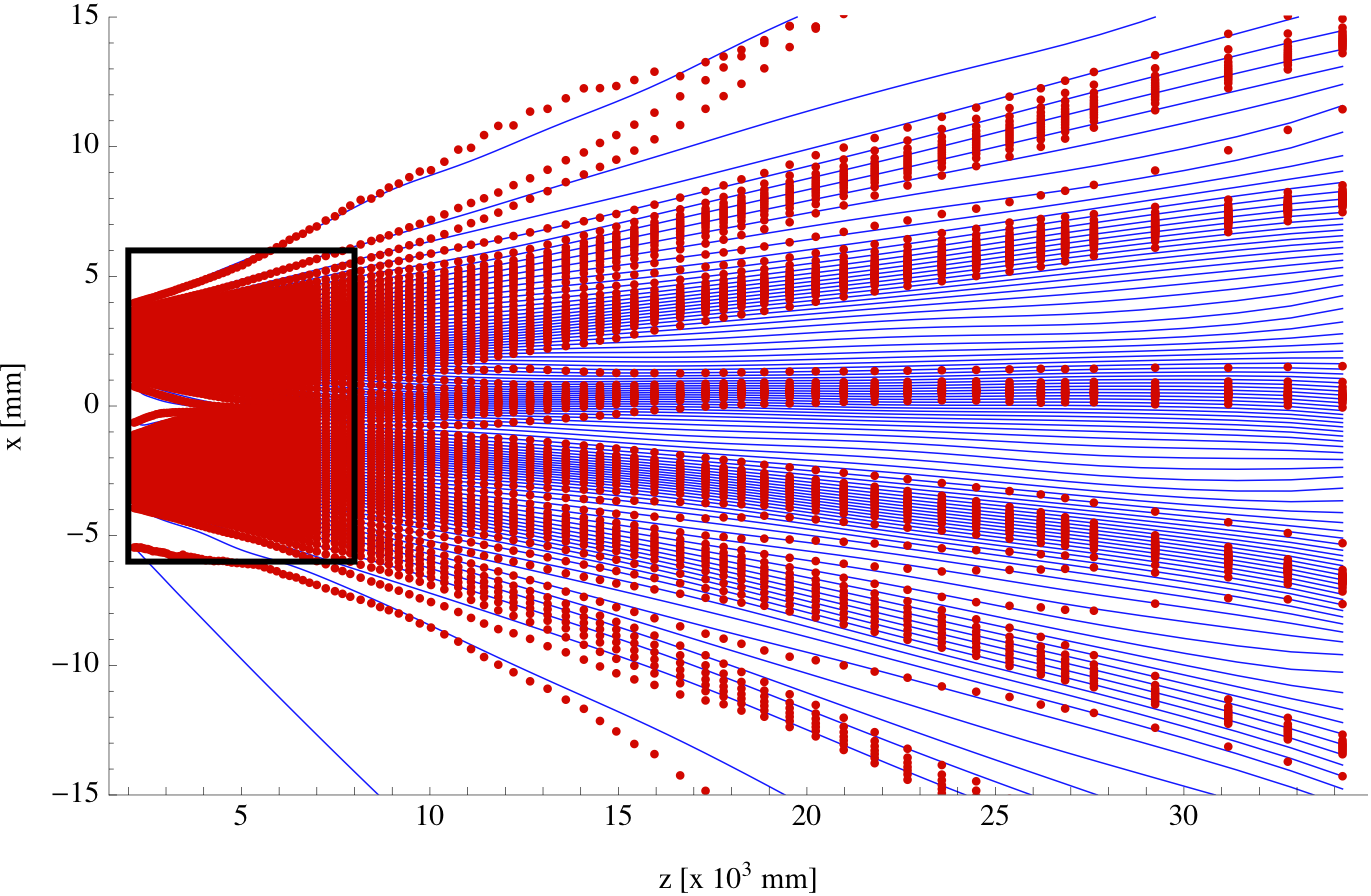}}
\centerline{(a)}
\centerline{\includegraphics{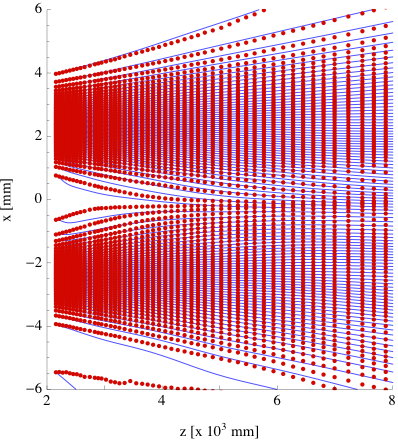}
\hspace{0.3in}\includegraphics{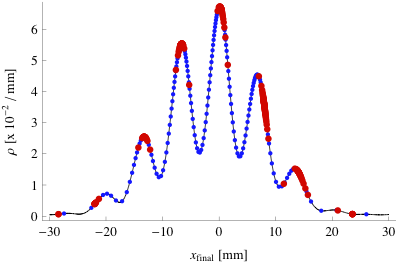}}
\centerline{(b)\hspace{3in}(c)}
\caption{Comparison of the average photon trajectories (red dots) and the Bohm trajectories (blue lines) after using density estimation on the collected experimental data. The inset (b) shows that the photon and Bohm trajectories now move similarly. Part (c) demonstrates that the photon trajectories congregate much better in the high probability regions.}
\end{figure}

Many adherents to Bohm's version of quantum mechanics assert that the trajectories are what particles {\it actually} do in nature. From the experimental results above no one would claim that photons actually traversed these trajectories, since the momentum was only measured on average and the pixel size of the CCD is still quite large. Other views of Bohm's trajectories do not go as far as to claim that they are what particles actually do in nature. But instead, the Bohm trajectories can be viewed simply as hydrodynamical trajectories \cite{Burghardt2001,Wyatt2005} that have equations of motion with an internal force that appears when one changes from a phase space to a position space discription. Recently, it has also been shown that the Bohm trajectories can in fact be generated {\it without any equations of motion} \cite{Coffey2010}. The probability density is represented at each time by a special {\it centroidal Voronoi tessellation}, and a mapping from one tessellation to the next produces trajectories that are identical to Bohm's trajectories in many cases. In this way, one concludes again that they Bohm trajectories are simply hydrodynamical and kinematically portraying the evolution of the probability density. The average photon trajectories can be viewed likewise.

This work was supported in part by a research grant from the Welch Foundation (grant number F-0362).

% Appendix
\section*{Appendix}

We found several minor errors while analyzing the Matlab codes provided to us by Kocsis {\it et al.} \cite{Kocsis2011}. Below is a listing of each error, which were contained in two files {\it Bohmdataread.m} and {\it DBAnalyze.m}. All numbers on the left refer to lines numbers of the original code. Some lines have been reformatted to fit on the page.

\subsection*{Bohmdataread.m}

{\ttfamily\small
\begin{verbatim}
69    Imgs(i,1).probh = Imgs(i,1).subh / sum(Imgs(i,1).subh);
70    Imgs(i,1).probv = Imgs(i,1).subv / sum(Imgs(i,1).subv);
\end{verbatim}
}
\noindent Lines 69--70 attempt to normalize each density image by dividing the count at each pixel by the total number of counts at all the pixels. This procedure, however, does not take into account the size of each pixel, and the fact that later in {\it DBAnalyze.m} (lines 210--225) each density image is rescaled or magnified so the pixel size changes. The correct procedure normalizes after each image has been magnified, and does a numerical integration over each image's range.

\subsection*{DBAnalyze.m}

{\ttfamily\small
\begin{verbatim}
323    %Recall that k_x(weak)/k = tan(sin^-1((V-H)/(H+V)*coeff))
        where coeff is
324    %the measurement coefficient, in units of rad/rad. 
...
349    fdata(i,1).imdata(j,1).kxkWeak = tan(asin(rmimag));
\end{verbatim}
}
\noindent Lines 323--349 compute the ratio of $k_x / |\mathbf{k}|$. In Kocsis {\it et al.} Equation~2 uses just the arcsin, with no mention of the tangent. We obtained better results with the removal of tangent function.

{\ttfamily\small
\begin{verbatim}
382    fdata(i,1).imdata(j,1).kxkWiseman(k,1) = tan(asin(Ltemp(j,2)));  
\end{verbatim}
}

\noindent In line 382 \texttt{Ltemp(j,2)} represents the slope $dx/dz$ of each probability conserved trajectory (Bohm). Based upon the geometry of the experiment, $\tan\theta = dx/dz$ where $\theta$ is the angle $\mathbf{k}$ makes with the $z$-axis. Similarily, $\sin\theta = k_x/|\mathbf{k}|$, and so $k_x/|\mathbf{k}| = \sin(\arctan(dx/dz))$. This discussion, however, is unnecessary since for trajectory building (see below) what is actually needed is simply $dx/dz$, so this line needs to be:

{\ttfamily\small
\begin{verbatim}
382'    fdata(i,1).imdata(j,1).kxkWiseman(k,1) = Ltemp(j,2);  
\end{verbatim}
}

{\ttfamily\small
\begin{verbatim}
415    cdfx = cdfx +
              (fdata(i,1).imdata(j+1,1).z-fdata(i,1).imdata(j,1).z) *
              (interp1(fdata(i,1).imdata(j,1).xreal,
                       fdata(i,1).imdata(j,1).kxkWeak,
                       cdfx,'cubic',0));
\end{verbatim}
}

\noindent Line 415 updates each photon trajectory with a new position, $x_{j} = x_{j-1} + \Delta z \cdot k_x/|\mathbf{k}|$, where the $k_x/|\mathbf{k}|$ is found by interpolation of the values obtained on the pixels. The correct trajectory position update should be $x_{j}=x_{j-1} + \Delta t \cdot k_x$. If we assume that $\Delta z = k_z \Delta t$, and $k^2 = k_x^2 + k_z^2$ (no $y$ dependence), then we can write the update in terms of $k_x/|\mathbf{k}|$ with, $x_{j} = x_{j-1} + \Delta z \cdot (|\mathbf{k}|/k_z)\cdot k_x/|\mathbf{k}|$, where $|\mathbf{k}|/k_z = (1-k_x^2/|\mathbf{k}|^2)^{-1/2}$.

{\ttfamily\small
\begin{verbatim}
416    cdfxWise = cdfxWise + 
                  (fdata(i,1).imdata(j+1,1).z-fdata(i,1).imdata(j,1).z)*
                  (interp1(fdata(i,1).imdata(j,1).CDFxrealset,
                           fdata(i,1).imdata(j,1).kxkWiseman,
                           cdfx,'cubic',0));
\end{verbatim}
}

\noindent Line 416 updates each Bohm trajectory with a new position. The \texttt{cdfx} in the \texttt{interp1()} function should be \texttt{cdfxWise}, so that each Bohm trajectory is correctly updated. In terms of $\Delta z$ this line is correct if the change for line 382 above is made.

% References

\bibliographystyle{apsrev4-1}

\bibliography{paper}

\end{document}